\documentclass{emulateapj}

\usepackage{graphics}      
\usepackage{amsmath}
\usepackage{amssymb}

\def\mj{M$_{\rm J}$}
\def\rj{R$_{\rm J}$}
\def\etal{{et~al.\,}}
\def\mo{M$_\odot$\ }
\def\ro{R$_\odot$}

\def\teff{$T_{\rm eff}\,$}
\def\teffs{$T_{\rm eff}$s$\,$}

\def\sles{\lower2pt\hbox{$\buildrel {\scriptstyle <}
   \over {\scriptstyle\sim}$}}
\def\sgreat{\lower2pt\hbox{$\buildrel {\scriptstyle >}
   \over {\scriptstyle\sim}$}}

\slugcomment{Accepted to Ap.J. May 3, 2011}

\begin{document}

\title{The Dependence of Brown Dwarf Radii on Atmospheric Metallicity and Clouds: Theory and Comparison with Observations}

\author{Adam Burrows\altaffilmark{1}, Kevin Heng\altaffilmark{2}, and Thane Nampaisarn\altaffilmark{1}} 

\altaffiltext{1}{Department of Astrophysical Sciences, 
Peyton Hall, Princeton University, Princeton, NJ 08544; burrows@astro.princeton.edu, tnampais@astro.princeton.edu}

\altaffiltext{2}{Institute for Advanced Study, School of Natural Sciences, 1 Einstein Drive, Princeton, NJ 08540, U.S.A.;
ETH Zurich, Institute for Astronomy, Wolfgang-Pauli-Strasse 27, CH-8093, Zurich, Switzerland;
kheng@phys.ethz.ch}

\begin{abstract}
Employing realistic and consistent atmosphere boundary
conditions, we have generated evolutionary
models for brown dwarfs and very-low-mass stars (VLMs)
for different atmospheric metallicities ([Fe/H]), with and without
clouds. We find that the spread in radius at a given mass
and age can be as large as $\sim$10\% to $\sim$25\%,
with higher-metallicity, higher-cloud-thickness
atmospheres resulting quite naturally in larger radii.
For each 0.1 dex increase in [Fe/H], radii
increase by $\sim$1\% to $\sim$2.5\%, depending upon age and mass.
We also find that, while for smaller masses and older ages
brown dwarf radii decrease with increasing
helium fraction ($Y$) (as expected), for more massive brown dwarfs
and a wide range of ages they increase with helium fraction.
The increase in radius in going from $Y=0.25$ to $Y=0.28$ can be as
large as $\sim$0.025 \rj\ ($\sim$2.5\%). Furthermore, we find
that for VLMs an increase in atmospheric metallicity from 0.0
to 0.5 dex, increases radii by $\sim$4\%, and from -0.5 to
0.5 dex by $\sim$10\%. Therefore, we suggest that opacity
due to higher metallicity might naturally account for the apparent
radius anomalies in some eclipsing VLM systems. Ten to twenty-five
percent variations in radius exceed errors stemming from
uncertainities in the equation of state alone.  This serves
to emphasize that transit and eclipse measurements
of brown dwarf radii constrain numerous effects collectively,
importantly including the atmosphere and condensate cloud models,
and not just the equation of state.  At all times, one is testing
a multi-parameter theory, and not a universal radius$-$mass
relation.
\end{abstract}



\section{Introduction}
\label{intro}

For the last fifteen years, the evolutionary, spectral, and atmospheric 
theory of brown dwarfs has evolved in tandem with the explosive growth 
in the number of known examples.  That number now stands near 1000.
This maturing theory has been used to interpret spectroscopic and 
photometric observations of such substellar-mass objects (SMOs) and, 
thereby, to derive effective temperatures (\teff), gravities ($g$), 
bolometric luminosities, compositions, masses ($M$) , and ages ($t$). The focus of 
study has been the new spectroscopic types, L and T, that have emerged
to accomodate objects with \teffs from $\sim$500 K to $\sim$2200 K. 
Both chemical equilibrium and non-equilibrium models have found 
roles and gradual improvements in molecular opacity and thermochemical 
databases suggest that objects with molecular atmospheres may someday 
be studied with the same precision as are stars with simpler atomic 
atmospheres, the latter albeit after almost a century of development.

However, most brown dwarfs are found as singletons, and those 
in binaries have only rarely been used to derive dynamical masses
with which to provide independent checks on 
spectroscopically-derived masses (Konopacky et al. 2010; Liu, 
Dupuy, \& Leggett 2010). Too often, however, such comparisons are
cast as fundamental tests of core assumptions in the models. Little mention is made of
the equal importance of atmospheric opacities, metallicities,
cloud models, and ages in determining whether theory
fits data.  In fact, these last-mentioned quantities are
crucial in determining the physical properites of a given 
brown dwarf. 

In addition to obtaining independent masses, measuring 
the radii of brown dwarfs can be an equally powerful
means of testing the associated physical theory 
(Burgasser, Burrows, \& Kirkpatrick 2006).  However, 
here too one is testing a multi-parameter theory, and not
a universal radius ($R$)$-$mass relation.  The latter exists
only for cold, very old, objects of a given elemental 
composition.  Hot, evolving brown dwarfs with a priori unknown 
atmospheric and interior compositions (and opacities) can span a (wide)
range (perhaps 25\%) of radii for a given mass and age. The brown dwarf's atmospheric 
opacities determine the rate with which heat escapes the convective core,
and at which the core entropy evolves, and it is the core entropy which 
sets the radius, for a given mass and internal composition. 
 
Gas-phase atmospheric opacities can be augmented by
silicate and iron cloud opacities.  The presence of clouds,
for which there is as yet no rigorous or credible model, 
complicates the theory and interpretation of the measured
physical properties of brown dwarfs. The thicker the atmosphere, 
the slower the radius shrinks and the larger is the radius at 
a given mass and age. Since the gas-phase opacities depend 
upon composition (atmospheric metallicity) and extant cloud models are 
still crude, without additional constraints on these factors 
there is an inherent ambiguity in the comparison between 
measured and theoretical radii that should temper any 
conclusions drawn. What is more, rapid rotation and (perhaps)
magnetic activity (Chabrier, Gallardo, \& Baraffe 2007; 
Morales et al. 2009) can further complicate the interpretation
of an observed radius and what one can deduce from differences between
measured and model radii.

For a given mass and age there is a physical, 
realistic {\it range} of possible radii that varies with 
atmospheric metallicity and cloud optical depth, both of which
would need to be measured to properly falsify theoretical models. For simplicity, 
evolutionary models have frequently been provided by theorists for only 
a restricted set of parameters (e.g., at solar metallicity) and for a single 
ad hoc prescription for clouds (or with no clouds) (Burrows et al. 
1993,1997,2001,2006; Chabrier \& Baraffe 1997; Baraffe et al. 1998,2003; Saumon \& Marley 
2008).  However, these models have been applied quite generally, with an assumption of 
universality that is not warranted. As a consequence, any deviation between
these theoretical models and physical measurements, be they of radius or mass,
must first account for the effects of different atmospheres before declaring
anything about, for example, the underlying equation of state or 
fundamental assumptions of the modeling exercise. Moreover, without
a good constraint on the age of the brown dwarf (perhaps from 
a good estimate of the primary's age and the assumption of coevality),
any radius measurement is of correspondingly limited utility if the 
goal is to test structural theory beyond its ``zeroth-order" aspects.
  
Recently, numerous exoplanet transit surveys have collectively discovered
and characterized more than one hundred transiting companions to nearby stars. 
These surveys include {\it Kepler} (Koch et al. 2010; Borucki et al. 2010), 
{\it CoRoT} (Baglin et al. 2006), WASP (Pollacco et al. 2004), 
OGLE (Udalski et al. 2002; Konacki et al. 2003), and HAT (Bakos et al. 2002).
Importantly, byproducts of such surveys, and their radial-velocity follow-up, are radii and 
masses for almost all of them, enabling the type of structural study to which we alluded above. 
Most of the transiting companions have masses near or below that of Jupiter
and most of them are very close to their primaries.  Such proximity
introduces the need to incorporate the effects of irradiation (and perhaps
tidal heating) in any interpretative study of the radius$-$mass relation
of objects in this mass range.  However, a small subset of the transiting 
companions found are quite massive (above $\sim$15 \mj) and the radii of 
such SMOs should not be affected by stellar irradiation (Burrows et al. 2007). 
Therefore, data for these more massive objects can be used to constrain brown dwarf 
theory. Of most interest for such a study are {\it Kepler's} LHS 6343 (Johnson et al. 2010),
CoRoT-15b (Bouchy et al. 2011), CoRoT-3b (Deleuil et al. 2008), and WASP-30b 
(Anderson et al. 2011). Since they reside in a physical region that is seamlessly 
connected to that of brown dwarfs, the very-low-mass stars (VLMs) OGLE-TR-122b (Pont et al. 2005)
and OGLE-TR-123b (Pont et al. 2006), though of stellar mass, test the same
basic theories.  

In this paper, we calculate a new suite of evolutionary models with new
sets of atmospheric boundary conditions for various atmospheric metallicities, with and
without clouds (i.e., clear).  With these models, we are able to demonstrate 
and quantify the dependence of the radius of a brown dwarf of a given mass and age
on the variety of different atmospheric characteristics expected. Brown dwarfs
should vary in metallicity and cloud cover.  Higher-metallicity atmospheres cool
more slowly and, therefore, result in larger brown dwarf radii, all else being equal.  Cloudy
atmospheres similarly retard radius shrinkage and lead to larger radii at a 
given mass and age.  Though we do not yet know how to incorporate this possibility,
higher-metallicity atmospheres may also have thicker cloud decks, amplifying
the effect of increasing metallicity.  Be that as it may, one should expect
a range of realistic brown dwarf radii at a given mass and age.
Our paper quantifies this range to highlight this important fact, 
while also exploring the dependence on the helium mass fraction ($Y$).

With the models we generate, we fit various of the recently measured transiting
brown dwarf radii.  We have chosen for this study the brown dwarfs LHS 6343C (Johnson et al. 2010), 
CoRoT-15b (Bouchy et al. 2011), CoRoT-3b (Deleuil et al.2008), and WASP-30b 
(Anderson et al. 2011), and the VLMs OGLE-TR-122b (Pont et al. 2005)
and OGLE-TR-123b (Pont et al. 2006), but the generic nature of the expected 
spread in radius at a given mass and age is the true focus of this investigation. 

While documenting representative theoretical differences in radius evolutions for 
different metallicities and helium fractions (with and without clouds), we at the same time
demonstrate that radius$-$mass measurements do not constrain solely  
the equation of state. The atmospheric opacities (and, less so, the helium 
fraction and overall internal compositions) are important factors in the 
interpretation of any radius measurement. Moreover, the significant 
remaining uncertainties in the ages of all these systems, and the difficulties 
and potential systematic inaccuracies in the measurements themselves, partially 
undermine any attempt to arrive at definitive conclusions concerning the 
accuracy or usefulness of any theoretical evolutionary model. Nevertheless, 
the magnitude of the expected theoretical spread in radii for brown dwarfs 
at a given mass and age should be more robust, and, as we demonstrate, is 
certainly not trivial.

In section \ref{data}, we summarize the data for the transiting objects
upon which we focus in this paper.  Then, in section \ref{method} we
review the techniques we have employed to calculate the evolutionary models, 
using an atmosphere/spectral code to obtain the boundary conditions 
for our Henyey evolutionary code.  In section \ref{general}, we 
discuss the general trends and behaviors of the full model suite, after which
in section \ref{individual} we determine individual object characteristics
from comparisons with the collection of new models.  Finally, in 
section \ref{conclusion}, we review the overarching conclusions 
concerning the radius$-$mass relation that have emerged from this study.

\section{Data Summary}
\label{data}

One of the most interesting objects in the observed set is LHS 6343C 
(Johnson et al. 2010), discovered using {\it Kepler}. It has a measured 
mass of $62.9 \pm 2.3$ \mj, a measured radius of $0.833 \pm 0.021$, 
and orbits one of a pair of M dwarfs (LHS 6343A, with an inferred mass of $\sim$0.37 M$_{\odot}$)
with a period of 12.71 days. From the minimal chromospheric activity, the authors 
suggest an age greater than 1 Gyr (gigayear), and prefer a range of $\sim$1 Gyr to $\sim$5 Gyr. 
Anderson et al. (2011) quote a metallicity ([Fe/H]) for LHS 6343A of $0.28\pm0.07$,
but Johnson et al. (2010) provide a value of $0.04 \pm 0.08$. The ``triple" nature of
this system complicated the extraction of physical parameters.  We quote
inferred metallicities for the primary of LHS 6343C (and for the 
other primaries discussed below) under the assumption that the 
primary and secondary metallicities of coeval objects formed
in the same context are likely to be similar and that the primary's [Fe/H] 
can serve as a guide when fitting secondary characteristics.

CoRoT-15b (Bouchy et al. 2011) has a measured mass of $63.3 \pm 4.1$ \mj, a measured 
radius of $1.12^{+0.30}_{-0.15}\;{\rm R}_\text{J}$, and orbits its F7V star 
primary with a period of 3.06 days.  Bouchy et al. (2011) suggest an age 
for CoRoT-15b of $\sim$1.14$-$3.35 Gyr and a metallicity of $\sim$$0.1\pm0.2$, basically 
solar.  Deleuil et al. (2008) obtain a mass for CoRoT-3b of $21.66 \pm 1.0\;{\rm M}_\text{J}$
and a radius of $1.01\pm0.07\;{\rm R}_\text{J}$.  These authors prefer an age and metallicity
for the F3V primary CoRoT-3 of $\sim$2 Gyr and $-0.02\pm0.06$, respectively, the latter 
consistent with solar. CoRoT-3b's orbital period is 4.26 days.

WASP-30b (Anderson et al. 2011) has a measured mass of $60.96 \pm 0.89\;{\rm M}_\text{J}$ 
and a measured radius of $0.89\pm0.021\;{\rm R}_\text{J}$. Anderson et al. (2011) quote a
metallicity for WASP-30 of $-0.08\pm0.10$, also basically solar, and prefer an age of $2.0 \pm 1.0$ Gyr.
WASP-30b is orbiting its F8V primary with a period of 4.16 days.

The very-low-mass stars (VLMs) OGLE-TR-122b and OGLE-TR-123b have
measured masses of $96 \pm 9\;{\rm M}_\text{J}$ (Pont et al. 2005)
and $89 \pm 12\;{\rm M}_\text{J}$ (Pont et al. 2006), respectively.
Their measured radii are $1.17^{+0.23}_{-0.13}\;{\rm R}_\text{J}$
and $1.30\pm0.11\;{\rm R}_\text{J}$. Pont et al. (2005,2006) suggest 
that the ages of these VLMs are probably less than $0.5$ Gyrs. The
metallicity of OGLE-TR-122 is estimated to be $0.15\pm0.36$, and, hence, 
can be considered unconstrained. OGLE-TR-122b is orbiting a solar-like star 
with a period of 7.3 days and OGLE-TR-123b is orbiting an F star with a 
period of 1.8 days. Both are at large distances ($\sim$1000 and $\sim$1600 
parsecs) and, hence, their primaries are difficult to characterize 
accurately.  An interesting eclipsing system is 2MASS J0535 (Stassun 
et al. 2006), for which, curiously, the least massive of the pair
has the higher measured \teff. However, though its masses are both in the brown dwarf
regime ($0.054\pm0.005$ M$_{\odot}$ and $0.034\pm0.003$ M$_{\odot}$) and the 
radii of these two objects are measured to be $0.669\pm0.034$ \ro and $0.511\pm0.026$ \ro,
respectively, the estimated system age is $\sim$1 Megayear (Myr).  At such young ages,
one needs a theory for their formation in the context of the protostellar disk. 
Extant evolutionary models for isolated spherical objects that assumed the 
initial conditions were lost on timescales of tens of Myrs are not suitable 
for comparisons at Myr ages.







\section{Methodology \& Assumptions}
\label{method}

Our spectral and atmosphere methods have previously been described in Burrows et al. 
(1993,1997,2003,2006), but we repeat the salient points here for 
completeness.  Given the surface gravity ($g$), effective temperature 
($T_{\rm eff}$) and cloud properties, the planar code \texttt{COOLTLUSTY} 
solves for the atmosphere and spectrum of an object using the hybrid 
technique of complete linearization and accelerated lambda iteration 
(Hubeny 1988; Hubeny \& Lanz 1995), adopting the solution to the (gray) Milne 
problem as a first guess.  Convection is treated using mixing length 
theory, where the mixing length $L_{\rm mix}$ in units of 
the pressure scale height ($H_p$) is set to 1.0.
Iteration proceeds until adjacent layers of the 1D 
atmosphere are in radiative equilibrium.  We have computed 
cloud-free evolutionary tracks for $L_{\rm mix} = 0.5 H_p$, $H_p$ and $2 H_p$
and find that within the ranges of effective temperature ($T_{\rm eff}=200$--3000 K) and
surface gravity ($\log_{10}{g} = 3.5$--5.5) considered in this paper, the models generated are
essentially identical. The main gas-phase species used in these theoretical atmospheres are H$_2$,
He, H$_2$O, CO, CH$_4$, N$_2$, NH$_3$, FeH, CrH, Na I, and K I and our
spectral model employs 30000 frequency points from $\sim$0.35 to 300 microns.
The opacities are taken from Sharp \& Burrows (2007). 

To generate molecular abundances in chemical equilibrium at a given atmospheric 
metallicity, we employed the thermochemistry found in Sharp \& Burrows (2007) and 
followed nearly 500 species (with over 150 condensates) containing 27 elements.  
For these abundance calculations, the atmospheric element fractions for a given metallicity
were taken from Allende-Prieto, Lambert, \& Asplund (2001),
Asplund et al. (2005), and Allende-Prieto \& Lambert (2005), which
replaced the Anders \& Grevesse (1989) compositions we used previously.

Refractory silicates condense out at temperatures below $\sim$2300 K and above $\sim$1700 K.
In our chemical code, these species are rained out (Burrows \& Sharp 1999) 
to determine the resulting gas-phase abundances. However, the clouds of the 
condensed refractories have a meteorology, physical extent, and particle-size distribution
that are as yet unconstrained (Ackerman \& Marley 2001; Helling et al. 2001,2004).

Therefore, to incorporate clouds into the spectral and atmosphere 
models, we introduce a flexible parametrization that allows us to vary 
cloud opacities and effects.  For this study, we assume that the numerous cloud 
species are mimicked by one representative species (in this case 
forsterite, Mg$_2$SiO$_4$) that extends from a base
at 2300 K to the intersection point of the atmospheric
temperature-pressure profile with the forsterite 
``Clausius-Clapeyron" condensation line. 

The baseline distribution of the cloud particle density is assumed
to follow the gas-phase pressure profile, but we multiply this baseline reference
distribution by a shape function, $f(P)$, to determine the actual model cloud
distribution in pressure space. $f(P)$ is never greater than one and can be made
to cut off sharply at the top and bottom of the cloud. Specifically, we define $f(P)$ as:
\begin{equation}
f(P) =
\begin{cases}
\left(P/P_{\rm u}\right)^{s_{\rm u}}, & P \le P_{\rm u} \\
f_{\rm cloud}, & P_{\rm u} \le P \le P_{\rm d} \\
\left(P/P_{\rm d}\right)^{-s_{\rm d}}, & P \ge P_{\rm d} \, , \\
\end{cases}
\label{eq:cloud}
\end{equation}
where $P_{\rm d}$ is the pressure at the cloud base and $P_{\rm u} < P_{\rm d}$.
Here, $P_{\rm u}$ is the pressure at the intersection of the atmospheric $T/P$ profile and the
forsterite condensation curve.  The indices ${s_{\rm u}}$ and ${s_{\rm d}}$ define
the rapidity with which the clouds are cut off on their upper and lower boundaries.
The larger they are, the sharper the cutoff. When $f_{\rm cloud} = 1$ the cloud has
a flat portion in its middle between $P_{\rm u}$ and $P_{\rm d}$.  Hence, the
parameters ${s_{\rm u}}$, ${s_{\rm d}}$, and $f_{\rm cloud}$ define the cloud spatial structure.

We chose $\sim$2300 K as our cloud base since it is
where the most refractory species (calcium-aluminum silicates)
condense. We chose the forsterite condensation line to cap this
region because forsterite has one of the lowest condensation temperatures
(for a given pressure). Hence, the physical region
where refractory clouds reside is more extensive than
where forsterite alone would reside. To attempt a multi-species
meteorological model, incorporating the optical properties
of all the expected refractories and the role of convection
in grain growth, and to pretend to model the various particle
sizes and shapes and the interaction between the various
grain species (mantling, collisions, etc.), would introduce
extra complexity wholely out of proportion with the sparse
dataset available for these very-remotely-sensed objects.

Note that if only forsterite clouds obtained,
the ``Clausius-Clapeyron" condensation line would be the
natural cloud base (and not the $\sim$2300 K line), but that convection and turbulence
would still likely extend the cloud upward to lower pressures.  The characteristic
scale of this is often assumed to scale with the pressure scale
height. This is the motivation for introducing the parameter ${s_{\rm u}}$, 
thereby capturing this physical extension above the forsterite condensation line.

For the cloud models in this study, we set $s_{\rm u}=2$ and $s_{\rm d}=10$ and 
use a modal particle size $a_0=30$ $\mu$m (Sudarksy et al. 2000), 
where the size distribution is given by:
\begin{equation}
\frac{dN}{da} \propto \left( \frac{a}{a_0} \right)^6 \exp{\left[ -6 \left(\frac{a}{a_0} \right) \right]}\, .
\end{equation}
We use Mie scattering theory with a table of complex indices of refraction
as a function of wavelength to calculate the absorption and scattering opacities for
the grains in the clouds. In summary, with this general cloud model we place 
clouds where they may be likely to reside, over the collective condensation region,
and then vary the modal particle size of forsterite grains to create a model set
with a wide range of possible cloud optical thicknesses.

To mimic the transformation of a young, cloudy object evolving into an old,
cloud-free object, and the L-to-T transition, we linearly combine cloudy (${\cal S}_{\rm cloudy}$)
and cloudfree (${\cal S}_{\rm cloudfree}$) spectra using the phenomenological prescription:
\begin{equation}
\begin{split}
&{\cal W} = \mbox{min}\left\{1, \left(T_{\rm eff}/T_0\right)^p \right\}, \\
&{\cal S}_{\rm hybrid} = {\cal W} {\cal S}_{\rm cloudy} + \left(1-{\cal W}\right) {\cal S}_{\rm cloudfree}.\\
\end{split}
\end{equation}
Thus, $T_0$ is the transitional temperature at and below which clouds in the
L dwarf atmosphere begin to both disperse and reside below the photosphere,
while the power law index $p$ describes the speed of the transition.
Effectively, our hybrid models have five parameters: $s_{\rm u}$, $s_d=10$, $a_0$, $T_0$ and $p$.
We are guided by empirical measurements of the
color-magnitude diagram to choose $T_0=1200$ K and $p=4$.  Although 
our approach is intrinsically different from $f_{\rm sed}$ 
parametrization of Saumon \& Marley (2008), the resulting 
cloud structures and temperature profiles are similar across
the L-to-T transition.

An evolutionary Henyey code is used to establish the mapping
between $\{T_{\rm eff},g\}$ and $\{M,t\}$ (mass-age) pairs.  With the equation of
state (EOS) for hydrogen/helium mixtures of (Saumon, Chabrier, \& Van Horn 1995) 
and specified values of the deuterium and helium fractions, 
the code solves the equations of stellar structure. 

We emphasize that the hydrogen/helium EOS of Saumon, Chabrier, 
\& Van Horn (1995) employs the volume addition law for such 
mixtures and does not have a rigorous means to incorporate 
heavy elements in a consistent fashion.  Hence, though 
the core and atmosphere will certainly have the same [Fe/H],
we do not include heavies in our EOS.  Currently, there does 
not exist a reliable EOS for more complicated mixtures.  If 
the matter were an ideal gas, then one could easily incorporate
any range of elemental abundance mixtures by calculating the 
appropriate mean molecular weight. However, the material
in SMOs experiences significant Coulomb interaction effects.
Indeed, at the high densities and low entropies encountered
in their cores, the Coulomb effects are important.  At higher densities,
degeneracy effects predominate, and the EOS, depending as it does on 
the electron fraction per baryon, again simplifies.
However, with brown dwarfs we are often in the intermediate regime
where Coulomb and degeneracy effects compete, an awkward regime
to treat well. At times, theorists have employed altered helium
fractions to treat heavies, boosting $Y$ by an ``appropriate" 
amount and then using the H/He EOS with the enhanced $Y$ 
(Spiegel, Burrows, \& Milsom 2011; Guillot et al. 2006). 
At solar metallicity, [Fe/H] = 0.0142 and the effect of neglecting
heavies in the EOS, while not large, is still interesting. However, 
the reader should be aware that the results we quote do not automatically
incorporate heavies in the core EOS and that the stopgap of
using a ``suitably" enhanced $Y$ to gauge their effects 
remains an option. If this be the case, then a $3\times$solar 
metallicity would be equivalent to a $\sim$0.04 augmentation
in $Y$, comparable to the range of $Y$ we explore in this paper.

The boundary condition between the atmosphere and the convective core 
is the entropy at the radiative-convective transition layer
of the planar atmosphere, which is an output of \texttt{COOLTLUSTY}.  By merging
the evolutionary and spectral computations in this manner, we ensure that our
spectral-evolutionary calculations are self-consistent.

For this investigation, we calculate models from 1 \mj\ to 0.15 M$_{\odot}$ 
($\sim$157 \mj) and focus on three atmospheric metallicities: [Fe/H]=$\pm$0.5 and 0.0 (solar).
Hence, our models are 1) clear with the three different metallicities
and 2) cloudy hybrid with [Fe/H]=0.0 and 0.5.  These five models, along with the heritage model
from Burrows et al. (1997), are used to fit the measured radii
of the brown dwarfs LHS 6343C, CoRoT-3b, CoRoT-15b, \& WASP-30b 
and the very-low-mass (VLM) stars OGLE-122b and OGLE-123b 
and to arrive at the various general conclusions.

\section{General Theoretical Behavior}
\label{general} 

In this section, we discuss the qualitative behavior of both the radius evolution with time
and the radius$-$mass isochrones as a function of the presence or absence of clouds, 
the atmospheric metallicity, and the helium fraction.  The importance of including clouds in evolutionary models was
previously emphasized by Saumon \& Marley (2008). Figure \ref{fig:evolution} depicts
the temporal evolution of the radii of theoretical brown dwarfs 
at four different masses ($0.055$, $0.060$, $0.065$,
and $0.070$ M$_{\odot}$, equivalent to $\sim$58, $\sim$63, $\sim$68, and
$\sim$73 \mj) for models with cloudy atmospheres at [Fe/H] = 0.0  and 0.5 and for 
cloud-free atmospheres at [Fe/H] = 0.0.  The cloud model incorporates the cloud 
prescription described in section \ref{method}.  Since the physics of cloud particles and 
structure are not yet understood, this cloud model should be viewed merely as representative of 
the effects of clouds. However, it is neither extreme in its properties, nor unrealistic,
and introduces the extra degree of atmospheric opacity expected of such clouds.  

As Fig. \ref{fig:evolution} indicates, before $\sim$1 Gyr brown dwarf radii
evolve very quickly, but later their shrinkage moderates significantly.
Importantly, however, this figure shows the radius hierarchy that emerges from this study.  Models with
higher atmospheric opacity, due either to higher metallicity or to the presence of thick
clouds, have larger radii at a given mass and age than those with lower opacity.  
The latter would be associated with lower metallicity and/or the absence of clouds.  
Importantly, the natural spread in theoretical radii at a given mass and age can
be significantly larger than the quoted formal error bars of the measured radii.
To demonstrate this, we have superposed onto this figure the data for the three brown dwarfs
in the mass range $\sim$60 to $\sim$70 \mj\ for which we have useful transit radii and masses 
(LHS 6343C, WASP-30b, and CoRoT-15b).  We caution that the formal error bars in measured 
radii may be trumped by systematic errors in the observational analysis and that the true
error in the data may be larger than published.  In turn, there are, no doubt, ``errors" 
in the theoretical models that spring from uncertainties in the gas-phase 
opacities, uncertainties in the the equation of state, elemental abundances 
patterns for a given metallicity that are different from what is assumed, and the 
many uncertainties in designing accurate cloud models.  Nevertheless, as Fig. \ref{fig:evolution}
demonstrates, for reasonable stellar/(brown dwarf) metallicities 
expected in the solar neighborhood, the radius spread at a given mass 
and age can be as large as $\sim$0.2 \rj. This exceeds any reasonable error stemming from
uncertainities in the equation of state alone and serves to emphasize that measurements
of brown dwarf radii constrain a collection of effects, importantly including the
atmosphere and condensate cloud models.  Without an independent measure of 
the atmospheric metallicity and constraints on the surface clouds, and a 
good age constraint, a measurement of a brown dwarf radius may be less useful 
than some might expect for testing the hydrogen/helium equation of state. 

One way to understand the effect of enhanced atmospheric opacity on the radius of an SMO
is by way of the consequent alteration in atmospheric temperature/pressure profiles
(for a given T$_{\rm eff}$ and gravity) with increasing atmospheric metallicity
or upon the introduction of clouds. Figure \ref{tpcloud} depicts three sets of 
atmospheric thermal profiles at T$_{\rm eff}$ = 1300 K and five gravities
from $\log_{10} g$ = 3.5 to $\log_{10} g$ = 5.5.  The three sets are cloud-free (red),
cloudy at solar metallicity (green), and cloudy at 3$\times$solar metallicity (blue).
As is clear from the figure, when comparing cloudy with cloud-free models we notice that
at a given temperature, the pressure is everywhere lower for the cloudy model.
This is also the case when comparing cloudy/solar with the cloudy/3$\times$solar models.
This indicates that the entropy at depth, the core entropy, is correspondingly higher
for the atmospheres with enhanced opacity due either to increased metallicity 
or the presence of clouds. Since at a given mass higher entropy 
translates into larger radii, Fig. \ref{tpcloud} neatly demonstrates the 
opacity/radius connection and our general thesis.

Figure \ref{burrows97} portrays isochrones from 0.5 to 5.0 Gyr in radius$-$mass space 
from 1 \mj\ to 120 \mj\ for the ``heritage" model published by Burrows et al. (1997). Superposed are the corresponding 
data for the six objects of this study. There is a slight peak near $\sim$4 M$_{\rm J}$.
At larger masses (in the ``brown dwarf regime"), radii decrease with increasing mass,
monotonically shrinking as they age.  Depending upon age, for masses around $\sim$60 \mj\ to $\sim$70 \mj,
the radius then begins to increase with mass.  This is a manifestation of the onset of
significant thermonuclear burning and the transition to a star. Hence, there is a minimum
radius for a given model and age.  From 1 Gyr to 5 Gyr that minimum radius ranges 
from $\sim$1.1 \rj\ to $\sim$0.75 \rj\ for the five-model suite studied in 
this paper and from $\sim$0.9 \rj\ to 0.77 \rj\ for the 
Burrows et al. (1997) models.  As Fig. \ref{burrows97} shows, OGLE-122b and OGLE-123b 
are clearly stars, though a comparison with these heritage models requires OGLE-123b
to be quite young.  However, Pont et al (2005,2006) suggest that the OGLE objects 
might indeed be younger than 0.5 Gyrs. The quoted age range of LHS 6343 is between one and 
five billion years (Johnson et al. 2010) and Fig. \ref{burrows97} implies that  
the Burrows et al. (1997) model set could fit this object. Anderson et al. (2011) suspect that WASP-30b is
younger than 1 Gyr, but given the inherent ambiguities in such an estimate 
Fig. \ref{burrows97} indicates that the Burrows et al. (1997) models 
might provide an adequate fit. The age of CoRoT-3 is not well constrained 
and its radius error bars are large, making a
comparison between theory and measurement an easy, though rather useless, exercise. 
However, with a suggested age range of $\sim$1.14$-$3.35 Gyr (Bouchy et al. 2011),
CoRoT-15b might be the only object among the six we are highlighting in this paper for
which a solution using the Burrows et al. (1997) models might be problematic.

Figure \ref{RvsM_clear_cloudy} compares radius$-$mass trajectories 
for cloudy-atmosphere models and clear-atmosphere models at four 
different ages (0.5, 1.0, 3.0, and 5.0 Gyr). The clear models
have atmospheric metallicities ([Fe/H]) of -0.5, 0.0, and +0.5, while the cloudy
models have atmospheric metallicities of 0.0 and +0.5. As Fig. \ref{RvsM_clear_cloudy} 
demonstrates, increasing the metallicity increases the radius, with or without clouds.
We note that atmospheres without clouds, or a cloud effect, are not expected for metallicities 
of solar and above.  Nevertheless, for clear models at early ages, the spread in 
radius for brown dwarfs with masses between $\sim$30 \mj\ and $\sim$70 
\mj\ and for the metallicity range of this study can be as much as $\sim$0.15 \rj. 
For the same clear model set at late ages, that same spread 
narrows between $\sim$55 \mj\ and $\sim$70 \mj\ to $\sim$0.1 \rj. For the cloud models,
at early ages the radii can range by $\sim$0.15 \rj\ in going from [Fe/H]=0.0 
to [Fe/H]=0.5, but that range tightens at the later ages to $\sim$0.05 \rj\ for 
the more massive brown dwarfs.  However, the spread in radius from clear 
models at [Fe/H] = -0.5 to cloudy models at [Fe/H] = 0.5 can be as large 
as $\sim$0.25 \rj\ at early ages, and as large as $\sim$0.1 \rj\ at late ages.
This implies that each increment in [Fe/H] of 0.1 can translate into
a radius change of $\sim$1$-$2.5 \%, depending upon age and mass.

In the stellar realm above $\sim$90 \mj, atmospheric 
temperatures are above the condensation temperatures of iron and
refractory silicates.  As a consequence, and 
as we see in Fig. \ref{RvsM_clear_cloudy}, the 
radii of stars depend only weakly on the presence or absence of
clouds.  However, their radii are still modest functions
of atmospheric metallicity. Increasing the metallicity 
of a VLM star from 0.0 to 0.5 increases the radius by $\sim$4\% and from -0.5 to
0.5 by $\sim$10\%.  The latter percentage is slightly above what is oft-quoted as the 
discrepancy in radius between measurement and theory in the VLM regime.  Though we do not 
in this investigation discuss models in the $\sim$0.2 \mo to 0.25 \mo band, the 
implication of the systematic behavior of Fig. \ref{RvsM_clear_cloudy} in the stellar realm
is that opacity due to higher metallicity can naturally account
for this apparent radius anomaly in, for instance, the CM Dra (Morales et al. 2009) 
and KOI-126 (Carter et al. 2011) systems. Neither rotation nor magnetic activity 
(Chabrier, Gallardo, \& Baraffe 2007) need be the only (nor the dominant) explanation. 
Summarizing, the effect of atmospheric metallicity on the radii of brown dwarfs and VLMs is straightforward
and natural, and the magnitude of the potential spread in radii for reasonable ranges in
metallicity in the solar neighborhood is comparable to published apparent discrepancies.

Figure \ref{helium} recapitulates some of the information on Fig. \ref{RvsM_clear_cloudy}, but 
highlights the helium fraction dependence of the radius$-$mass relation and is for only solar 
atmospheric metallicity. Shown are trajectories for $Y = 0.25$ (blue/aqua) and $Y = 0.28$ (red/magenta), 
where $Y$ is the helium mass fraction.  At zero temperature, the higher the mean molecular weight
and the lower the electron-to-baryon ratio (both consequences of higher $Y$), the smaller
the brown dwarf.  At later ages and lower masses, Fig. \ref{helium} clearly demonstrates 
this tendency, and this is the widely-expected result.  However, for the 
higher temperatures and entropies which persist for more massive brown dwarfs,
the opposite is true $-$ a higher helium fraction leads to a larger radius.  The magnitude of the effect
is slight, but can amount to as much as $\sim$0.025 \rj.  As Fig. \ref{helium} shows, 
this slight ``reversal" continues into the stellar realm.  We note that the Burrows et al. (1997)
models (also shown on the figure for comparison) were calculated for $Y = 0.25$, but 
that a value of $Y=0.28$ is generally preferred (Saumon \& Marley 2008).

This slightly counterintuitive result at larger masses
is due to the enhanced core hydrogen burning rate for the 
larger $Y$, despite the slightly smaller hydrogen fraction.
In hydrostatic equilibrium, the fact that at higher mean molecular 
weight the adiabat is higher results in a higher core temperature
for an isentropic core.  Hence, larger $Y$ (which translates 
into a higher mean molecular weight) results in higher central 
temperatures, which generate higher thermonuclear powers. This also 
results in higher core entropies for a given convective mass and age, 
and, hence, larger outer radii. When the nuclear luminosity is very 
small, the larger $Y$ results in more compact configurations, as
is expected from the behavior of the EOS, and smaller outer
radii.  However, for larger masses, the core temperatures
eventually become significant, even if the object is not
destined to settle onto the main sequence. The condition
for the latter is that the core thermonuclear power equal
the atmospheric luminosity $-$ the first mass for which this
condition is satisfied defines the main sequence edge. However,
even for lower masses, thermonuclear power can retard the decay
of the core entropy, which for a given mass determines the outer
radius. That the specific heat decreases with increasing $Y$
is a slightly mitigating effect, but this is trumped by
the enhancement in nuclear burning rate. The same phemonenon
was identified in Spiegel, Burrows, \& Milsom (2011) in the
context of deuterium burning.  

Figure \ref{ratio} portrays the ratio of the nuclear 
luminosity to the surface photon luminosity for 
representative masses from 0.05 to 0.08 M$_{\odot}$ and 
$Y$ = \{0.25, 0.28\}.  When this ratio is equal to one we are on 
the main sequence. As Fig. \ref{ratio} demonstrates, even 
for masses near 0.05 M$_{\odot}$ and far below the main sequence, 
core burning is both non-trivial and larger for larger $Y$.
The higher nuclear luminosity for higher $Y$ also helps explain
why VLM radii are larger for larger $Y$ (Fig. \ref{helium}).

We end this section by reiterating that our core EOS does not 
automatically incorporate the slight effect of metallicity, 
but that one can approximately account for heavy elements in the EOS
by augmenting the helium fraction accordingly. This suggests, 
given the findings summarized in Fig. \ref{helium}, that 
properly including them in the EOS would further (though only slightly) 
increase the radii of VLMs and of the more massive brown dwarfs younger than $\sim$5 Gyrs, 
emphasizing yet again the potential role of metallicity in influencing
SMO radii.

\section{Models for Individual Objects}
\label{individual}

The six objects (LHS 6343C, WASP-30b, CoRoT-15b, CoRoT-3b, OGLE-TR-122b, and OGLE-TR-123b)
that we investigate in this paper have measured transit radii and radial-velocity 
masses.  The errors in the measured radii quoted in the discovery papers are usually purely 
statistical, yet systematic errors may actually dominate.   
Therefore, the reader should keep this in mind when we 
derive approximate physical properties for these objects from a comparison with our 
theoretical models.  There may in fact be more interpretive latitude than we have allowed 
ourselves. Given this caveat, we now proceed to discuss each individual brown dwarf or VLM.

\subsection{LHS 6343C}
\label{lhs6343_sec}


On Fig. \ref{lhs6343}, we have plotted four panels in radius$-$mass space 
from 40 \mj\ to 80 \mj, each for a different model set, with and without clouds and for [Fe/H]
equal to either 0.0 or 0.5. Four isochorones at 0.5, 1.0, 3.0, and 5.0 Gyr are
given on each panel.  The data from Johnson et al. (2010) for 
LHS 6343C ($M = 62.9 \pm 2.3$ \mj; $R = 0.833\pm0.021$ \rj) are superposed, where 
the errors are 1-$\sigma$.  Johnson et al. (2010) suggest an age range from 1 to 5 Gyrs.
LHS 6343C is seen to be a small brown dwarf, and, therefore, Fig. \ref{lhs6343} 
suggests that older ages, clearer models, and lower metallicities might
be preferred.  However, a range of combinations can be shown to fit.  For the cloudy
model, the older solar-metallicity models may be indicated (bottom right panel).  Anderson et al. (2011)
suggest a metallicty of $0.28\pm0.07$, but Johnson et al. (2010) quote $0.04 \pm 0.08$.
If we were to take the Johnson et al. number and note that low gas-phase metallicity may
suggest thinner clouds, we get a consistent set of characteristics in the lower-right-hand panel
with an age near 5 Gyrs.  Clear models at the higher metallicity fit for an age near $\sim$4$-$5 Gyrs.
Clear models at the lower metallitiy fit for ages greater than $\sim$2 Gyrs (bottom left panel).  
However, the  high-metallicity, cloudy model (top right panel) does not fit unless the age is greater 
than $\sim$7 Gyrs.  Curiously, the Burrows et al. (1997) models (see Fig. \ref{burrows97}) fit well for 
a suggested age of $\sim$2 Gyrs.

\subsection{WASP-30b}
\label{wasp30_sec}

Figure \ref{wasp30} is the same as Fig. \ref{lhs6343}, but for WASP-30b and with 
the [Fe/H] = 0.5 and [Fe/H] = 0.0 models switched (top $\leftrightarrow$ bottom). Superposed on all panels
is the data point for WASP-30b at $60.96 \pm 0.89\;{\rm M}_\text{J}$ and
$0.89\pm0.021\;{\rm R}_\text{J}$ (Anderson et al. 2011).  The metallicity of 
WASP-30 is quoted to be [Fe/H] = $-0.08\pm0.10$, consistent with solar. As Fig. 
\ref{wasp30} indicates, a variety of models and age$-$metallicity combinations fit the WASP-30b data.
Clear models with [Fe/H] = 0.0 fit well for $\sim$1$-$2 Gyrs.  Clear models with [Fe/H] = 0.5
fit well for ages from $\sim$2 to $\sim$3 Gyrs.  Cloudy models with [Fe/H] = 0.0 fit
well for ages of $3.0\pm1.0$ Gyrs and our cloudy model with [Fe/H] = 0.5 still fits near
ages of $\sim$5 Gyrs. Cloudy models with $Y = 0.25$ (not shown) fit at slightly younger ages.
As Fig. \ref{burrows97} suggests, the heritage models from Burrows et al. (1997)
fit WASP-30b for an age near $\sim$1 Gyr. Therefore, as Figs. \ref{wasp30} and \ref{burrows97}
together demonstrate, the WASP-30b radius is rather easily fit within its suggested mass, 
age, and metallicity constraints (soft as they are).

\subsection{CoRoT-15b}
\label{corot15_sec}

We superpose the data from Bouchy et al. (2011) for this 
brown dwarf ($62.9 \pm 2.3\;{\rm M}_\text{J}$; $1.12^{+0.30}_{-0.15}\;{\rm R}_\text{J}$) 
on Fig. \ref{corot15}, which is in the same format as Fig. \ref{wasp30}, but has a slightly
different range for the radius and mass axes. CoRoT-15b has the widest 
error bars for a measured radius among the set of brown dwarfs
upon which we focus in this study. Bouchy et al. (2011)
estimate a metallicity of $0.1\pm0.2$ (consistent with solar) and an age
for its primary in the range $\sim$1.14$-$3.35 Gyr.  As Fig. \ref{corot15} 
suggests, our solar-metallicity models fit in the lower age range to 
within $\sim$1-$\sigma$ to $\sim$1.5-$\sigma$, with the best fit 
for the [Fe/H] = 0.0 cloudy model.  However, our [Fe/H] = 0.5 models fit the suggested age range better,
with the clear [Fe/H] = 0.5 models fitting an age of $\sim$1 Gyr to within $\sim$1-$\sigma$
and the cloudy [Fe/H] = 0.5 models fitting anywhere in the suggested age range.
Cloudy models with [Fe/H] = 0.0 and clear models with [Fe/H] = 0.5 fit CoRoT-15b
almost equally well. Figure \ref{burrows97} indicates that the Burrows et al. (1997)
solar-metallicity models would fit only for very young ages less than $\sim$0.5 Gyrs.
If the ``highish" value of the radius survives, a high-opacity
atmosphere model, either due to high metallicity or thick clouds or both, may be 
indicated for CoRoT-15b.  However, the range of possible models available to fit this
object remains comfortably broad, particularly for ages younger than $\sim$2 Gyrs.

\subsection{CoRoT-3b}
\label{corot3_sec}

Figure \ref{corot3} is similar to Figs. \ref{lhs6343}, \ref{wasp30}, and \ref{corot15}, 
but is constructed for CoRoT-3b and has a lower mass range between 5 and 40 \mj.  
Deleuil et al. (2008) measure the mass of CoRoT-3b to be 
$21.66 \pm 1.0\;{\rm M}_\text{J}$ and its radius to be $1.01\pm0.07\;{\rm R}_\text{J}$.  CoRoT-3
has an estimated metallicity of $-0.02\pm0.06$ and, hence, the models
with solar metallicity may best describe the data.  Note that due to
the large uncertainty in its radius, many models, both
clear and cloudy with a variety of age$-$metallicity pairs, can fit CoRoT-3b
comfortably. Deleuil et al.(2008) quote an age near $\sim$2 Gyr.  Fortunately or unfortunately,
due to the large radius error bars, all the model sets plotted (cloudy and clear, 
[Fe/H] = 0.0 and 0.5) fit the CoRoT-3b data reasonably well. Higher 
metallicity/cloudy models favor older ages and low metallicity/clear models
favor younger ages, but ages between 1 and 5 Gyr are consistent with the quoted
mass/radius data for the putative estimated age and metallicity.
In addition, as Fig. \ref{burrows97} demonstrates, the Burrows et al. (1997) models
fit for any age younger than $\sim$6 Gyrs.

\subsection{OGLE-TR-122b and OGLE-TR-123b}
\label{ogle_sec}

Figure \ref{ogle} compares the measured radius$-$mass points for the VLM stars
OGLE-TR-122b and OGLE-TR-123b with cloudy and clear models at ages of from 0.5 to 
5.0 Gyrs and for the [Fe/H] = 0.0 and 0.5. Pont et al. (2005,2006) suggest that 
the ages of these VLMs are probably less than $0.5$ Gyrs. If this 
is the case, as Fig. \ref{ogle} suggests, most of our models provide acceptable fits, with the 
best fits for OGLE-TR-123b given by either the cloudy or the clear 
models with [Fe/H] = 0.5.\footnote{Recall that in the putative  mass regime 
of these stars, the atmospheric temperatures are near or above the silicate 
and iron condensation temperatures. As a result, the differences between 
cloudy and clear models are minimal.} Higher metallicities would fit OGLE-TR-123b 
even better, but all models fit OGLE-TR-122b. The large radius error bars, as 
well as the ambiguity in the age of the systems, limit the strength of any
general conclusions.  Moreover, given the difficulty of these measurements,
one should not discount the possibility that these radii might be revised 
in the future. Not shown on Fig. \ref{ogle} are models for [Fe/H] = -0.5 (see 
Fig. \ref{RvsM_clear_cloudy}), which predict slightly smaller radii at a given mass and age.
Nevertheless, whatever the age of these OGLE stars, and in this stellar mass realm,
the set of different models (all reasonable in their character and inputs) 
we have calculated allow a spread in expected radii of $\sim$10\%.

\section{Discussion and Conclusions}
\label{conclusion}

In this paper, we have generated a collection of evolutionary
models for brown dwarfs and very-low-mass stars for different atmospheric metallicities,
with and without clouds. These models employ realistic atmosphere boundary conditions
that allow us to consistently predict, given a detailed opacity model,
the evolution of the object's radius. We have sought to demonstrate 
that with transit or eclipse radius measurements one is testing a
multi-parameter theory, and not a universal radius$-$mass 
relation.  The spread in radius at a given mass 
can be as large as $\sim$0.1 \rj\ to $\sim$0.25 \rj (or $\sim$10\% 
to $\sim$25\%), with higher-metallicity, higher-cloud-thickness 
atmospheres resulting quite naturally in larger radii, all else 
being equal. For each 0.1 dex increase in atmospheric [Fe/H], the radius 
is expected to increase by $\sim$1\% to $\sim$2.5\%, 
depending upon age and mass. Therefore, in order to constrain 
the hydrogen-helium equation of state one must control for 
the metallicity and cloud model. If the goal is to test structural 
theory and the viability of a suite of theoretical models, absent 
measurements of, for example, the metallicity, and a good constraint 
on the age, any radius measurement is of correspondingly limited utility.
Conversely, one should expect a range of radii for the natural 
range of metallicities and possible cloud properties expected for 
substellar objects and VLM stars in the solar neighborhood.

In addition, we have calculated the effect of helium fraction on brown
dwarf and VLM radii and find that, while for smaller masses and older ages
radius decreases with increasing helium fraction (as expected), for more 
massive brown dwarfs and a wide range of ages it increases with  
helium fraction. This runs counter to common lore, which expects 
that higher mean molecular weights universally result in smaller radii.  
We find that the increase in radius in going from $Y=0.25$ to $Y=0.28$ 
can be as large as $\sim$0.025 \rj ($\sim$2.5\%). Furthermore, we suggest that
properly including the trace of heavy elements in the core EOS should
further augment this effect.

We do not suggest that the cloud model we constructed for this investigation 
is definitive, nor uniquely applicable.  Rather, we engineered a cloud 
model, and its corresponding opacities and optical depths, to be 
representative of the generic effect of the clouds we know 
reside in brown dwarf atmospheres.  We have sought merely to 
demonstrate that the presence  or absence of clouds has an important 
effect on the radius of a brown dwarf. 
%
%
We note that many other cloud models can be constructed
which may prove in the long run to be more viable.  Nevertheless, the qualitative
effect of clouds that we have highlighted is robust.  Similarly, and more straightforwardly,
atmospheric metallicity has a direct and clear effect on brown dwarf radii that
needs to be accounted for in any interpretation of radius measurements.

Ten to twenty-five percent variations in radius exceed any reasonable error stemming from
uncertainities in the equation of state alone and serve to emphasize that measurements
of brown dwarf radii constrain a collection of effects, importantly including the
atmosphere and condensate cloud models.  Without an independent measure of
the atmospheric metallicity, constraints on the surface clouds, and a
good age estimate, a measurement of a brown dwarf radius may be more 
difficult to interpret than previously thought.

Increasing the atmospheric metallicity of the VLM stars we studied in this paper from 0.0 to 0.5 
increases their radii by $\sim$4\%.  If we increase their atmospheric metallicitiy from -0.5 to
0.5, their radii increase by $\sim$10\%. The latter percentage is slightly above what is 
often quoted as the discrepancy in radius between measurement and theory in the VLM regime 
(Morales et al. 2009; Carter et al. 2011).  Though we do not in this paper 
discuss models in the $\sim$0.2 \mo\ to 0.25 \mo\ band, the clear
implication of the systematic behavior we have derived in the stellar realm
is that opacity due to higher metallicity might naturally account
for the apparent radius anomalies in some eclipsing VLM systems.

Be that as it may, the effect of metallicity and clouds on the radii of 
brown dwarfs and VLMs is straightforward and natural.  Therefore, 
we suggest it is necessary to incorporate these extra degrees 
of freedom into any interpretation of brown dwarf and VLM radius 
measurements and into any attempts usefully to constrain their 
equation of state.

\acknowledgments

We acknowledge useful conversations with Dave Spiegel and Nikku Madhudsudhan,
and Ivan Hubeny for his general support of the COOLTLUSTY code.
We also acknowledge support in part under NASA ATP grant
NNX07AG80G, HST grants HST-GO-12181.04-A and HST-GO-12314.03-A, and
JPL/Spitzer Agreements 1417122, 1348668, 1371432, and 1377197.

{}

\clearpage

\newcommand{\myscale}{0.5}
\begin{figure}[htp]			
\begin{center}
\includegraphics[scale=\myscale]{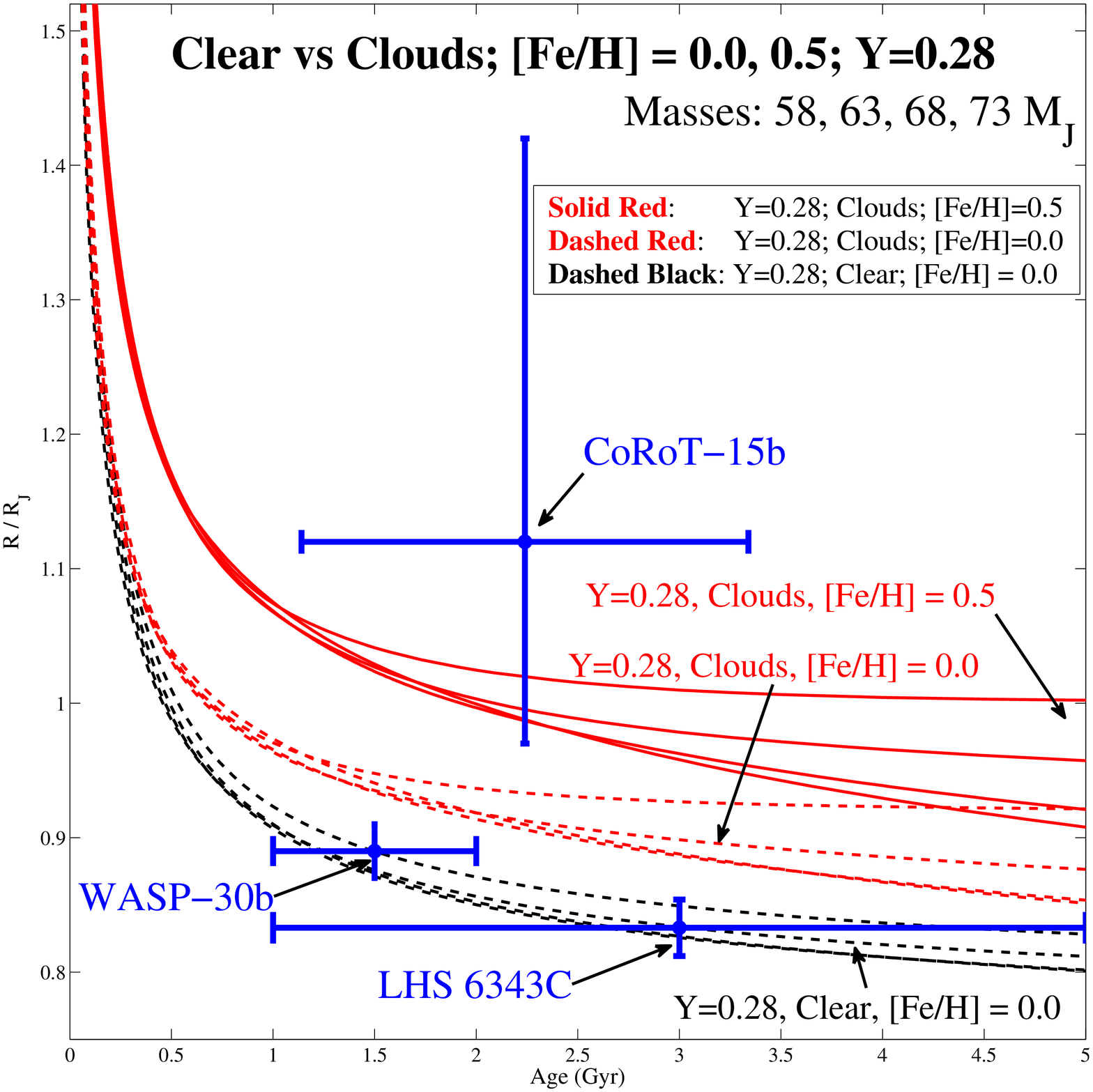} 
\end{center}
\caption{The evolution with age (in Gyrs) of the radii (in units of Jupiter's radius) of brown dwarfs
for representative models with 1) cloudy atmospheres (at [Fe/H] = 0.0 [dashed] and 0.5 [solid], shown in red)
and 2) clear atmospheres (at [Fe/H] = 0.0 [dashed], shown in black). (See section \ref{method} for a discussion of 
the cloud prescription.) The four masses shown are $0.055$, $0.060$, $0.065$,  
and $0.070$ solar mass (or approximately $58$, $63$, $68$, and 
$73$ Jupiter masses).  The helium mass fraction ($Y$) for each model 
is here set to $0.28$, i.e., roughly the solar value. This 
figure shows that the radius of a brown dwarf 
is an increasing function of metallicity and is larger for models
with clouds, all else being equal.  Three of the brown 
dwarfs which we highlight in this paper (LHS 6343C, WASP-30b, 
and CoRoT-15b) have masses within this range and are included (in blue) on the figure, 
with putative observational error bars. The formal errors in the age are 
the most uncertain. As this figure indicates, before $\sim$1 Gyr, the radius 
evolves quickly, but afterwards significantly decelerates its shrinkage.}
\label{fig:evolution}
\end{figure}
		
\clearpage
		
\begin{figure}[htp]			
\begin{center}
\includegraphics[angle=-90,width=6.5in,scale=\myscale]{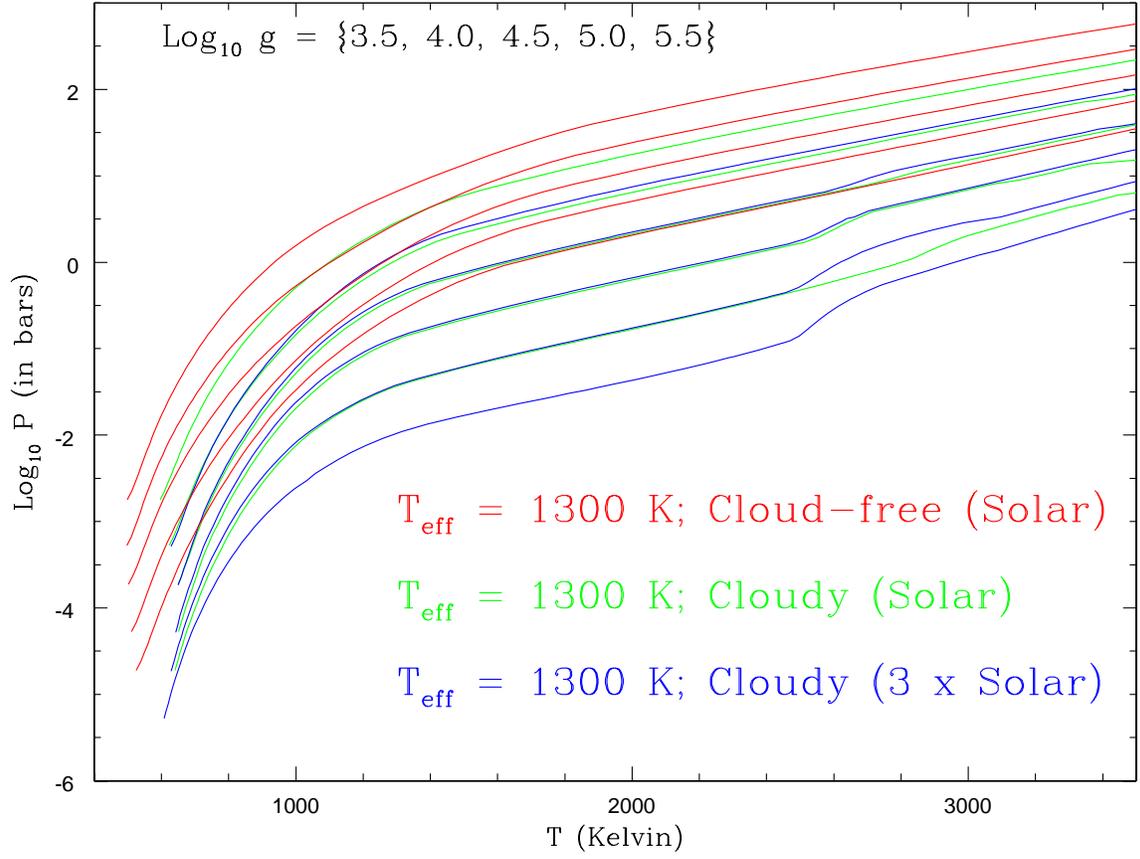} 
\end{center}
\caption{This figure depicts three sets of
atmospheric thermal profiles for T$_{\rm eff}$ = 1300 K 
and five gravities from $\log_{10} g$ = 3.5 to $\log_{10} g$ = 5.5. 
The three sets are cloud-free (red), cloudy at solar metallicity 
(green), and cloudy at 3$\times$solar metallicity (blue).
The pressure is in bars and is logarithmically displayed and the 
temperature is in Kelvin. At given T$_{\rm eff}$ and for a given 
model set, the higher gravity profiles are those with higher pressures.
See text in section 4 for a discussion of the import and meaning of this figure.}
\label{tpcloud}
\end{figure}
		
\clearpage

\begin{figure}[htp]			
\begin{center}
\includegraphics[scale=\myscale]{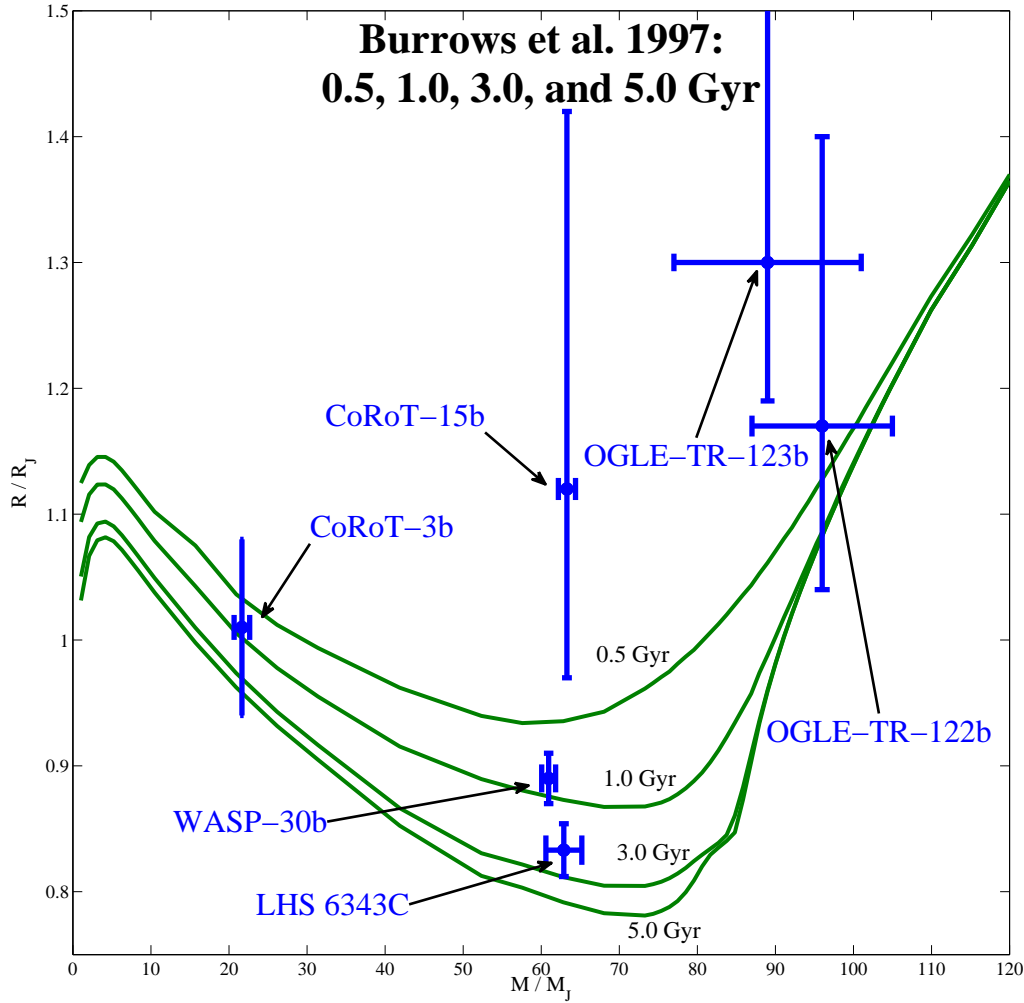} 
\end{center}
\caption{Radius (in units of Jupiter's radius) versus Mass (in Jupiter masses)
using the Burrows et al. (1997) ``heritage models" at ages $0.5$, $1.0$, 
$3.0$, and $5.0$ Gyr. Notice the slight peak near $\sim$4 M$_{\rm J}$,
the decrease in radius with increasing mass in the ``brown dwarf regime" at 
greater masses, and the monotonic radius shrinkage with increasing age.  
The curves start to rise near the main-sequence edge (near 
$\sim$75 M$_{\rm J}$) and OGLE-122b and OGLE-123b are clearly stars.
Six objects (four brown dwarfs, two VLMs) are shown in blue 
(with putative mass and age error bars): LHS 6343C (Johnson et al. 2010), WASP-30b 
(Anderson et al. 2011), CoRoT-3b (Deleuil et al. 2008), CoRoT-15b 
(Bouchy et al. 2011), OGLE-TR-122b (Pont et al. 2005), and 
OGLE-TR-123b (Pont et al. 2006). The quoted age of LHS 6343 
is between one and five billion years. WASP-30b is suspected to be 
younger (see Fig.\ref{fig:evolution}). Pont et al (2005,2006) suggest 
that the OGLE objects might be younger than 0.5 Gyrs, and, hence,
their measured radii might be consistent with these models.
The age of CoRoT-3 is not well constrained. Therefore, with the suggested age 
range of $\sim$1.14$-$3.35 Gyr for the star CoRoT-15 (Bouchy 
et al. 2011), CoRoT-15b is the only object in this set for
which a solution using these heritage models may be 
problematic. See the text for a discussion.} 
\label{burrows97}
\end{figure}
		
\clearpage
		
\begin{figure}[htp]			
\begin{center}
\includegraphics[scale=\myscale]{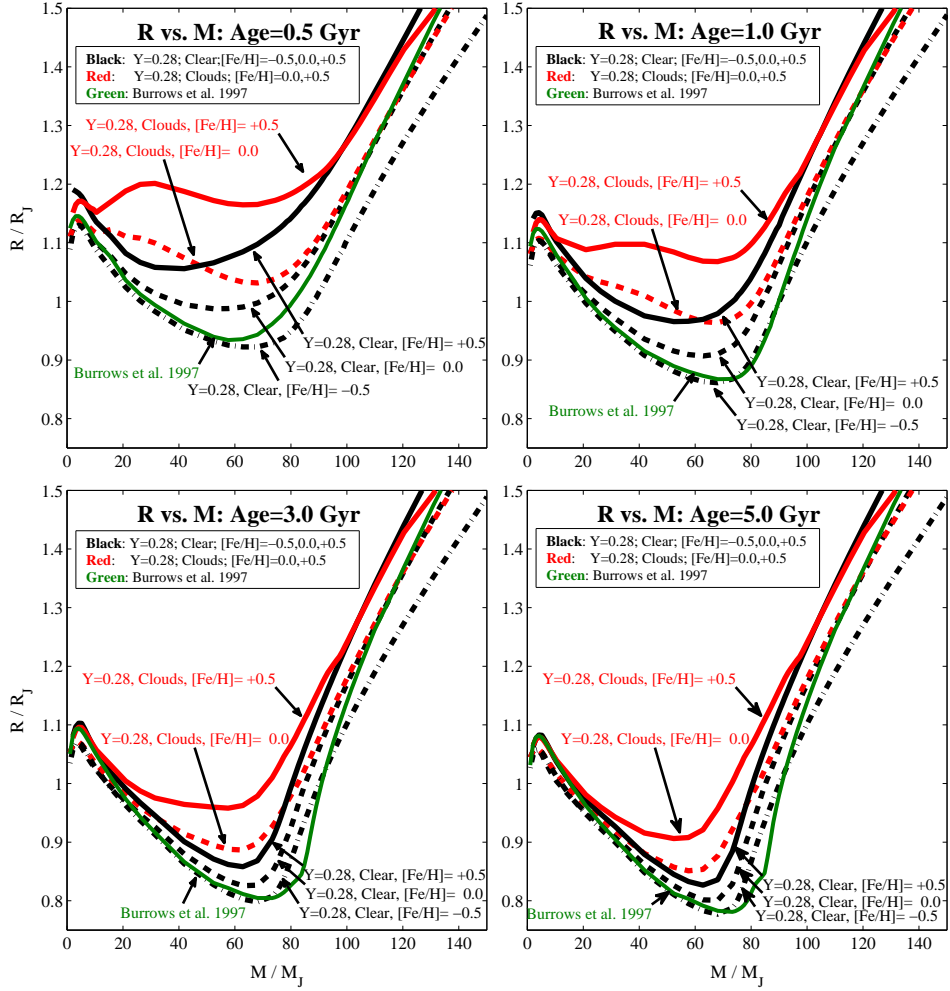} 
\end{center}
\caption{Plots of radius (in Jupiter radii) versus mass (in Jupiter masses) 
at ages of 0.5 (top left), 1.0 (top right), 3.0 (bottom left), and 5.0 
(bottom right) Gyr for several new models, all with $Y=0.28$.  The heritage models from Burrows et al. (1997)
(in green) are included for comparison. Cloudy-atmosphere models are shown in red and 
clear-atmosphere models are shown in black.  The clear-atmosphere models 
have metallicities ([Fe/H]) of -0.5, 0.0, and +0.5, while the cloudy-atmosphere 
models have metallicities of 0.0 and +0.5. A general trend is that  
the presence of clouds retards shrinkage, as does higher metallicity.
The model with [Fe/H] = 0.5 and a cloudy atmosphere generally
has the largest radii and the variation in radii among the models shown
can be 0.1$-$0.25 R$_{\rm J}$ at a given mass and age.}
\label{RvsM_clear_cloudy}
\end{figure}
		
\clearpage
		

\begin{figure}[htp]			
\begin{center}
\includegraphics[scale=\myscale]{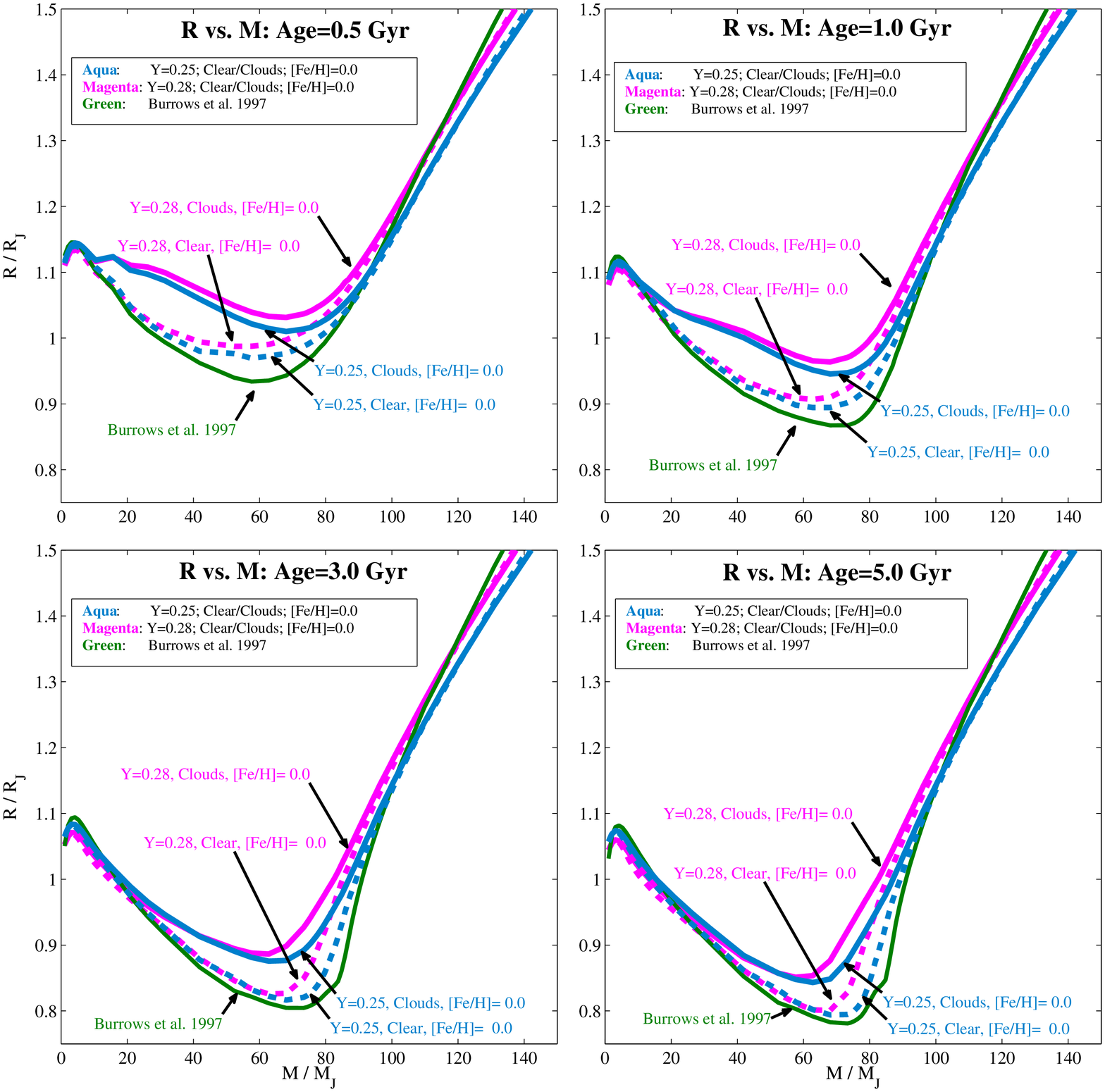} 
\end{center}
\caption{Similar to Fig. \ref{RvsM_clear_cloudy}, but highlighting 
the differences between models with two different helium fractions, 
$Y=0.25$ (blue/aqua) and $Y=0.28$ (magenta).  The gas-phase metallicity is solar
for all models shown. As a reference, the model from Burrows et al. 
(1997) is also plotted (in green). The solid lines 
represent models with clouds, while the dashed lines are 
those with clear atmospheres. The solid (cloudy) models are generally 
larger than the corresponding dashed (clear) models. Interestingly, 
the magenta ($Y=0.28$) models are larger than the blue/aqua ($Y=0.25$) models
in the mass range above $\sim$55-60 \mj\ at later ages (3.0 and 5.0 Gyr) 
and above $\sim$35-40 \mj\ at earlier ages (0.5 and 1.0 Gyr).  This is counter to
common lore, which suggests that planets with a higher molecular weight and
lower electron fraction should be smaller.  This is true only for cold planets.
Note that the Burrows et al. (1997) models were calculated for $Y=0.25$ 
(Saumon \& Marley 2008). See text for a discussion.}
\label{helium}
\end{figure}
	
\clearpage

\begin{figure}[htp]			
\begin{center}
\includegraphics[angle=-90,width=6.5in,scale=\myscale]{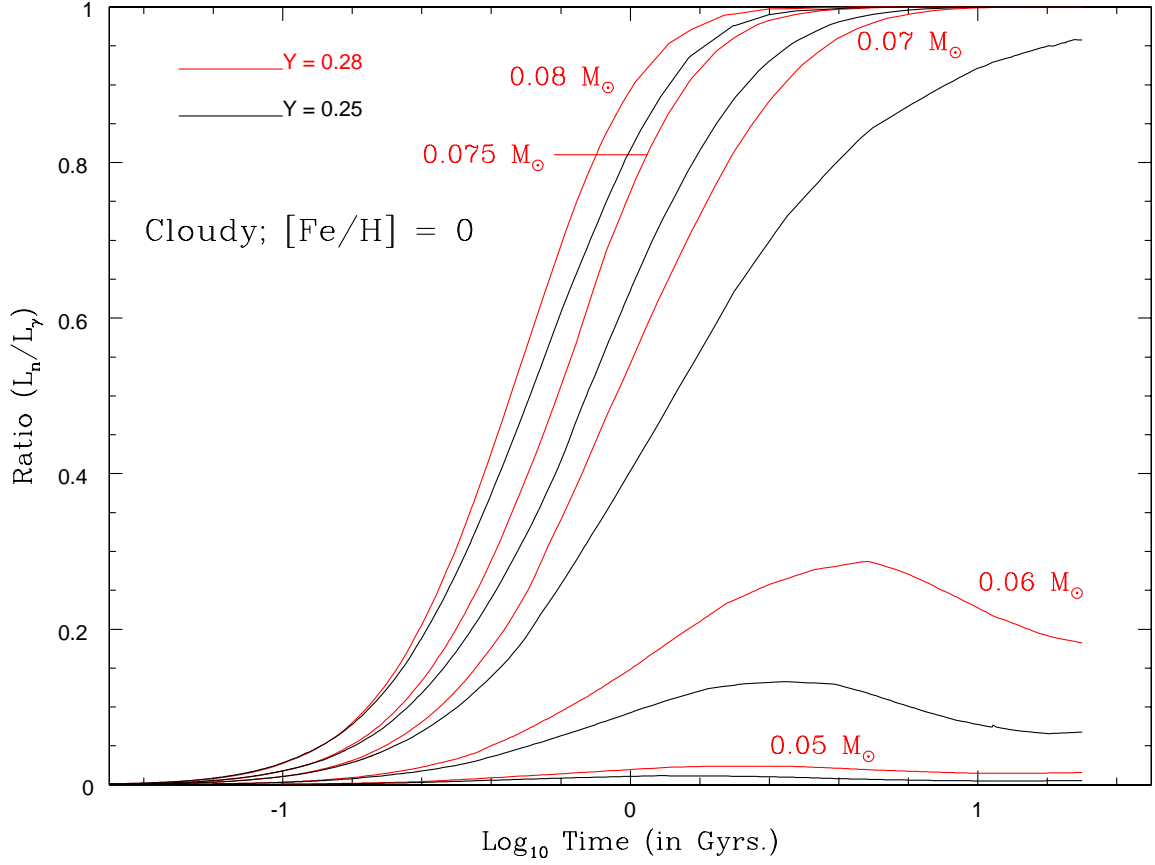} 
\end{center}
\caption{The ratio of the core nuclear
power to the surface photon luminosity versus age/time 
(in Gyrs.) for masses of 0.05, 0.06, 0.07, 0.075, and 0.08 M$_{\odot}$.  
The main sequence obtains when this ratio is one.
The red lines are for $Y$ = 0.28 and the black lines are for 
$Y$ = 0.25.  The mass of a given model is printed nearest 
the relevant $Y$ = 0.28 (red) line. The corresponding $Y$ = 0.25 (black) 
line is always below the $Y$ = 0.28 line. The solar-metallicity 
cloudy atmosphere model was used. As this figure demonstrates, even
for masses near 0.05 M$_{\odot}$ and far below the main sequence 
edge, core burning is both non-trivial and larger for larger $Y$.
The fact that the nuclear power is higher for higher $Y$ also helps explain
why VLM radii are larger for larger $Y$ (refer to Fig. 
\ref{helium}). See text for a discussion.}
\label{ratio}
\end{figure}
		
\clearpage 
	
\begin{figure}[htp]			
\begin{center}
\includegraphics[scale=\myscale]{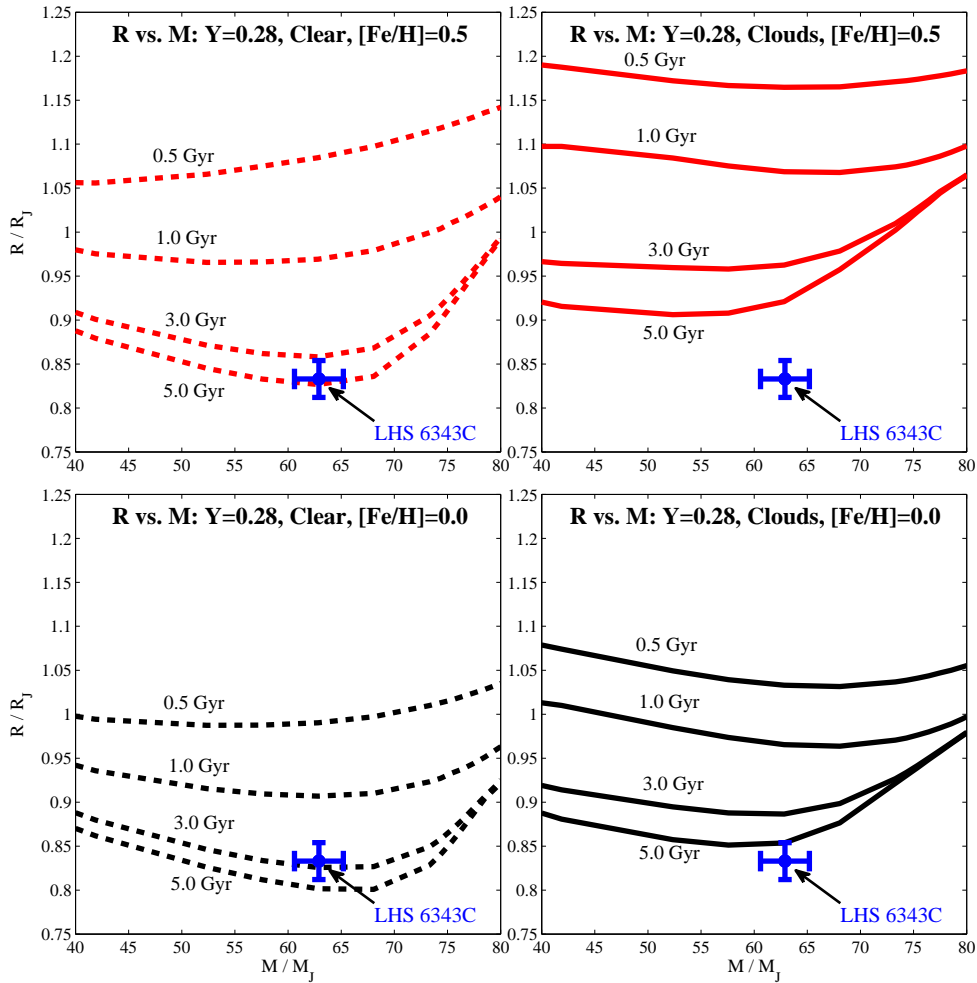} 
\end{center}
\caption{LHS 6343C Study: Radius (in units of Jupiter's radius) versus mass (in units of Jupiter's mass)
for clear models (left panels) and cloudy models (right panels) for [Fe/H] = 0.5 (top two panels)
and [Fe.H] = 0.0 (bottom two panels).  All the models shown are for a helium fraction, $Y$, of 0.28.
On each panel are shown isochrones at 0.5, 1.0, 3.0, and 5.0 Gyrs.  Superposed on all panels 
is the data point for LHS 6343C at $62.9 \pm 2.3\;{\rm M}_\text{J}$ and 
$0.833\pm0.021\;{\rm R}_\text{J}$ (Johnson et al. 2010).  The 
metallicity of LHS 6343C's primary, LHS 6343A, is suggested by Anderson et al. (2011) to
be [Fe/H] = $0.28\pm0.07$, i.e. super-solar, but Johnson et al. (2010) quote  
a value of $0.04 \pm 0.08$.  Clear models for [Fe/H] = 0.0 (solar) and [Fe/H] = 0.5
metallicities fit for ages greater than $\sim$2 Gyrs, 
but the cloudy model with $Y = 0.28$ and [Fe/H] = 0.0 also fits for late ages.
Models with $Y = 0.25$ fit slightly better, but given the systematic observational 
uncertainties, nothing substantive can be said about $Y$ for LHS 6343C.  The new model 
with clouds and super-solar metallicity does not fit for ages less than $\sim$7 Gyr.
As Fig.\ref{burrows97} demonstrates, the Burrows et al. (1997) models fit well for the 
suggested age of $\sim$2 Gyrs. See the text for a discussion of the 
implications of all these model comparisons.}
\label{lhs6343}
\end{figure}
		
\clearpage 
	
\begin{figure}[htp]			
\begin{center}
\includegraphics[scale=\myscale]{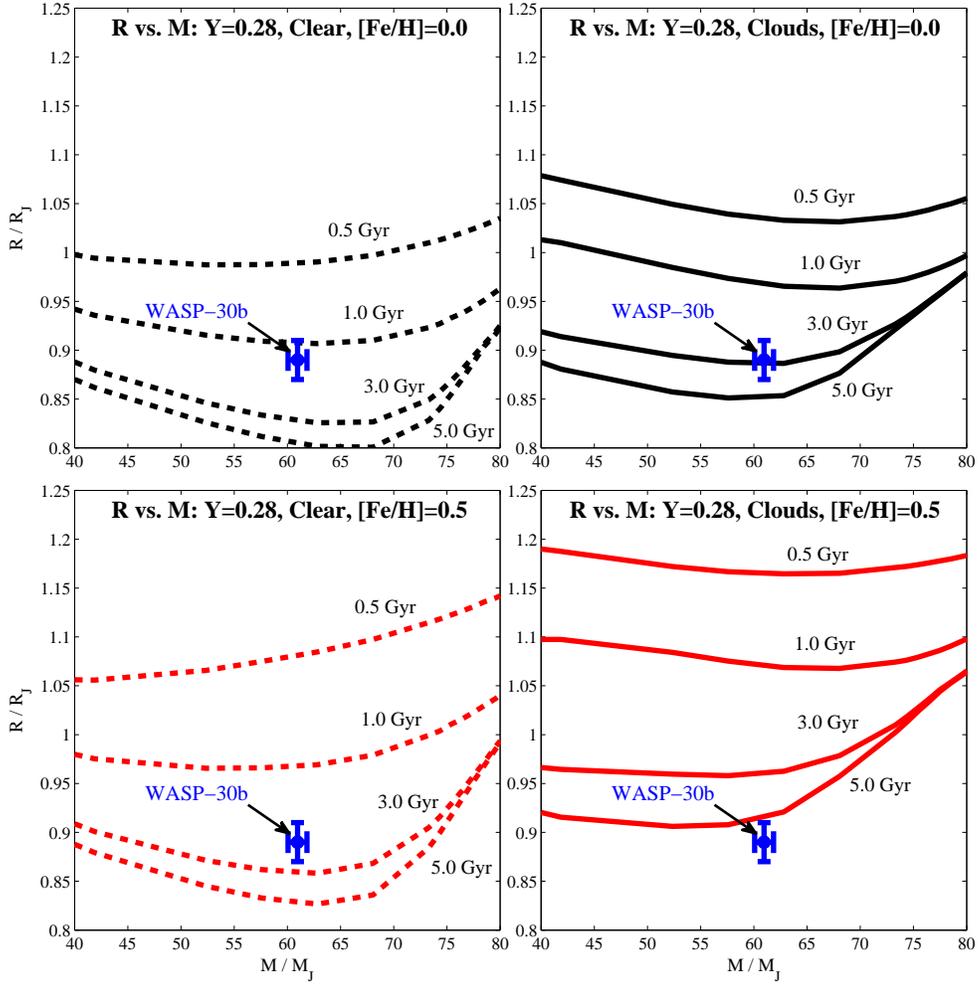} 
\end{center}
\caption{Similar to Fig. \ref{lhs6343}, but for WASP-30b and with the [Fe/H] = 0.5 models at the bottom
and the [Fe/H] = 0.0 models at the top. (Note that a slightly different range of radii on the ordinate is used.)
As in Fig. \ref{lhs6343}, all the models shown are for a helium fraction of 0.28
and isochrones at 0.5, 1.0, 3.0, and 5.0 Gyrs are plotted.  Superposed on all panels
is the data point for WASP-30b at $60.96 \pm 0.89\;{\rm M}_\text{J}$ and
$0.89\pm0.021\;{\rm R}_\text{J}$ (Anderson et al. 2011).  The
metallicity of WASP-30 is quoted to be [Fe/H] = $-0.08\pm0.10$, basically solar.
A variety of models and age$-$metallicity combinations fit the WASP-30b data.
Clear models with [Fe/H] = 0.0 fit well for $\sim$1$-$2 Gyrs.  Clear models with [Fe/H] = 0.5
fit well for ages from $\sim$2 to $\sim$3 Gyrs.  Cloudy models with [Fe/H] = 0.0 fit
well for ages of $3.0\pm1.0$ Gyrs and our cloudy model with [Fe/H] = 0.5 still fits near 
ages of $\sim$5 Gyrs. Cloudy models with $Y = 0.25$ fit at slightly younger ages.
As Fig. \ref{burrows97} suggests, the heritage models from Burrows et al. (1997)
fit WASP-30b for an age near $\sim$1 Gyr. See the text for a discussion of these 
fits and conclusions concerning WASP-30b.}
\label{wasp30}
\end{figure}
		
\clearpage
		
\begin{figure}[htp]			
\begin{center}
\includegraphics[scale=\myscale]{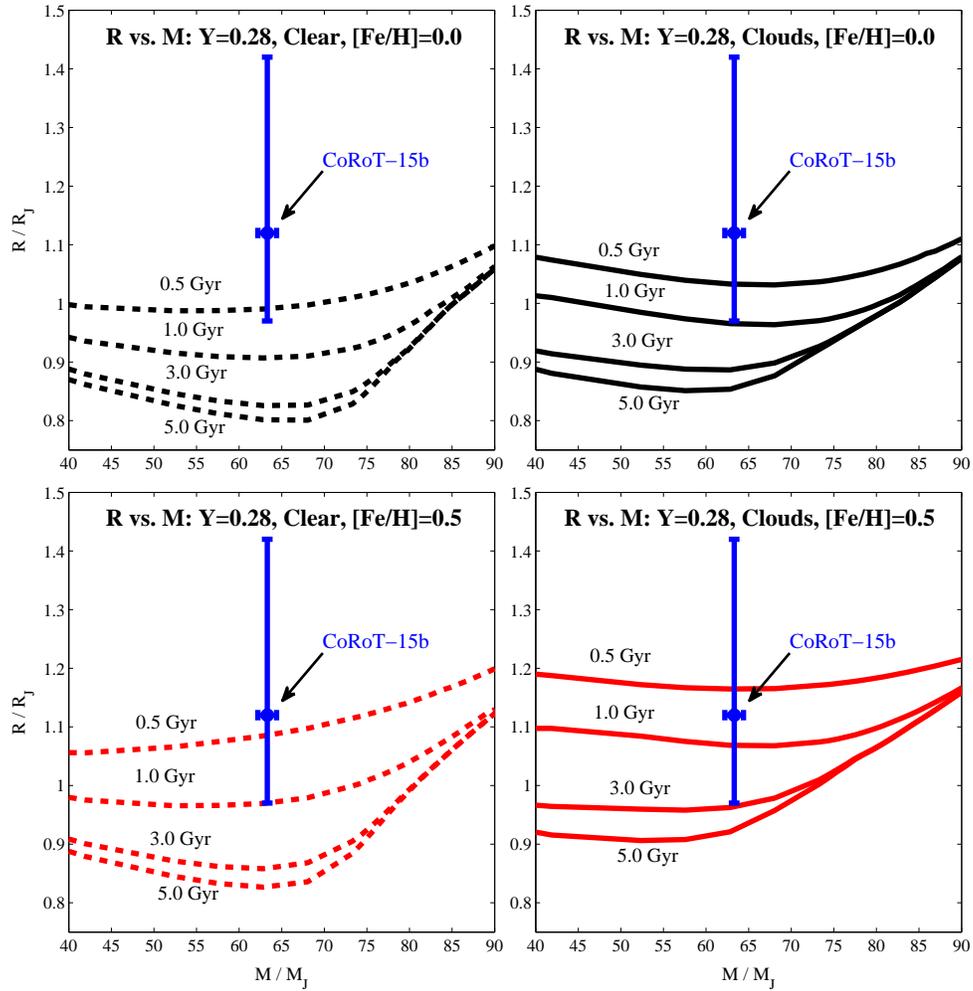} 
\end{center}
\caption{Same as Fig. \ref{wasp30}, but for CoRoT-15b and with slightly 
different ranges for the radius and mass axes. The mass of CoRoT-15b is 
measured to be $62.9 \pm 2.3\;{\rm M}_\text{J}$ and its radius is 
measured to be $1.12^{+0.30}_{-0.15}\;{\rm R}_\text{J}$ (Bouchy et al. 2011).  CoRoT-15 
has an estimated metallicity of $0.1\pm0.2$, again basically solar.  
Bouchy et al. (2011) suggest that its parent star has an age 
in the range $\sim$1.14$-$3.35 Gyr. Our solar-metallicity ([Fe/H] = 0.0) models can fit 
the lower age range to $\sim$1-$\sigma$ to $\sim$1.5-$\sigma$, with the best fit for the [Fe/H] = 0.0 
cloudy model.  However, our [Fe/H] = 0.5 models fit the suggested age range better,
with the clear [Fe/H] = 0.5 models fitting an age of $\sim$1 Gyr within $\sim$1-$\sigma$
and the cloudy [Fe/H] = 0.5 models fitting anywhere in the suggested age range.
Cloudy models with [Fe/H] = 0.0 and clear models with [Fe/H] = 0.5 fit CoRoT-15b
almost equally well. Figure \ref{burrows97} suggests that the Burrows et al. (1997) 
solar-metallicity models would fit only for very young ages less than $\sim$0.5 Gyrs. 
See the text for details.}
\label{corot15}
\end{figure}
\clearpage
		
\begin{figure}[htp]
\begin{center}
\includegraphics[scale=\myscale]{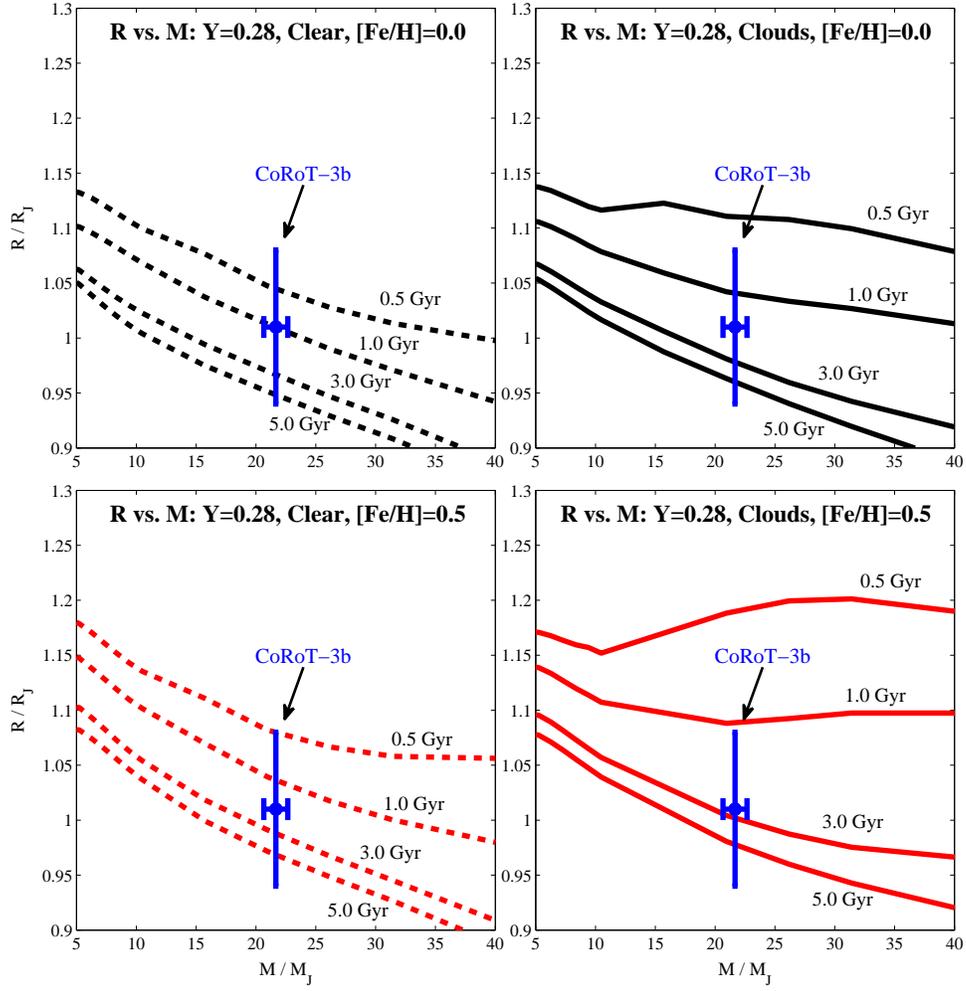}
\end{center}
\caption{Similar to Figs. \ref{lhs6343}, \ref{wasp30}, and \ref{corot15}, but for CoRoT-3b
and for a lower brown-dwarf mass range between 5 and 40 \mj.  The clear models are on 
the left panels and the cloudy models are on the right panels.  
The top models are for [Fe/H] = 0.0 and the bottom models are 
for [Fe/H] = 0.5. The measured mass of CoRoT-3b is $21.66 \pm 1.0\;{\rm M}_\text{J}$ 
and its measured radius is $1.01\pm0.07\;{\rm R}_\text{J}$ (Deleuil et al. 2008).  
See text for a discussion.}
\label{corot3}
\end{figure}
\clearpage

\begin{figure}[htp]			
\begin{center}
\includegraphics[scale=\myscale]{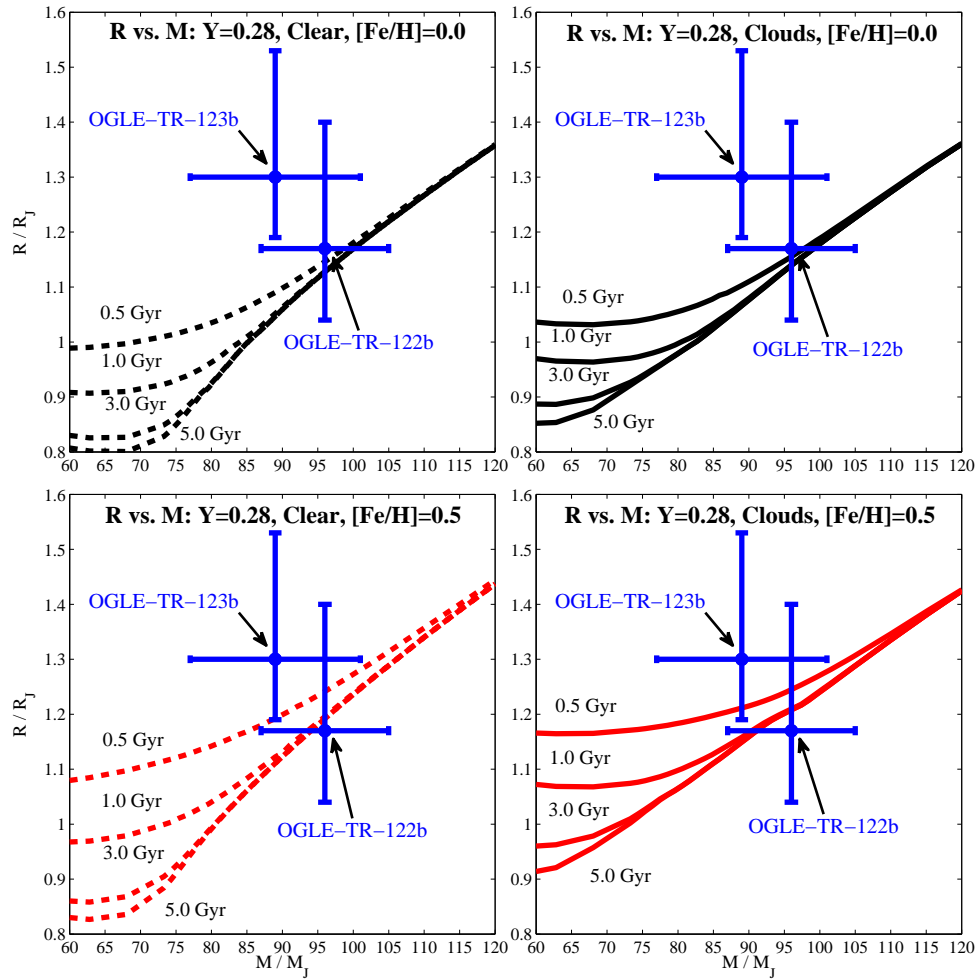} 
\end{center}
\caption{Similar to the four previous four-panel figures, the corresponding 
theoretical radius (in \rj) versus mass (in \mj) model set for OGLE-TR-122b and OGLE-TR-123b.  Four isochrones 
at 0.5, 1.0, 3.0, and 5.0 Gyrs are provided for the [Fe/H] = 0.0 (top panels) 
and 0.5 (bottom panels) and for clear (left panels) and cloudy (right panels) models. 
The data for OGLE-TR-122b and OGLE-TR-123b are taken from 
Pont et al. (2005) and Pont et al. (2006), respectively. Since 
these objects have measured masses of $96 \pm 9\;{\rm M}_\text{J}$ 
(OGLE-TR-122b) and $89 \pm 12\;{\rm M}_\text{J}$ (OGLE-TR-123b), 
the plots are for a mass range from 60 to 120 \mj. The measured 
radii are $1.17^{+0.23}_{-0.13}\;{\rm R}_\text{J}$ and 
$1.30\pm0.11\;{\rm R}_\text{J}$ for OGLE-TR-122b and OGLE-TR-123b,
respectively, and the plotted radius range is 0.8 to 1.6 \rj. 
See text for a discussion of the issues involved.}
\label{ogle}
\end{figure}

\end{document}